%% file: mu_jamming.tex
\documentclass[conference]{IEEEtran}
\IEEEoverridecommandlockouts
\usepackage{cite}
\usepackage{amsmath,amssymb,amsfonts}
\usepackage{algorithmic}
\usepackage{graphicx}
\usepackage{textcomp}
\usepackage{xcolor}
\usepackage{hyperref}
\usepackage{pgfplots, tikz}
\pgfplotsset{compat=1.12}

\usepackage{amsthm, bm, siunitx}
\DeclareSIUnit{\belmilliwatt}{Bm}
\DeclareSIUnit{\dBm}{\deci\belmilliwatt}

\input{sections/preamble.tex}

\def\BibTeX{{\rm B\kern-.05em{\sc i\kern-.025em b}\kern-.08em
    T\kern-.1667em\lower.7ex\hbox{E}\kern-.125emX}}

\IEEEoverridecommandlockouts
\IEEEpubid{\makebox[\columnwidth]{978-1-6654-0579-9/22/\$31.00~\copyright{}2023 IEEE \hfill} \hspace{\columnsep}\makebox[\columnwidth]{ }}

\begin{document}

\title{Sensing-Assisted Receivers for Resilient-By-Design 6G MU-MIMO Uplink
\thanks{
The authors acknowledge the financial support by the Federal Ministry of Education and Research of Germany in the programme of “Souverän. Digital. Vernetzt.”. 
Joint project 6G-life, project identification number: 16KISK002.
This work was also supported by the Bavarian Ministry of Economic Affairs,
Regional Development and Energy within the project 6G Future Lab Bavaria.
}
}

\author{\IEEEauthorblockN{Vlad C. Andrei\IEEEauthorrefmark{1},
Xinyang Li\IEEEauthorrefmark{2}, Ullrich J. M\"{o}nich\IEEEauthorrefmark{3} and
Holger Boche\IEEEauthorrefmark{4}}
\IEEEauthorblockA{\textit{Chair of Theoretical Information Technology} \\
\textit{Technical University of Munich}\\
Munich, Germany \\
Email: \IEEEauthorrefmark{1}vlad.andrei@tum.de,
\IEEEauthorrefmark{2}xinyang.li@tum.de,
\IEEEauthorrefmark{3}moenich@tum.de,
\IEEEauthorrefmark{4}boche@tum.de}}
\maketitle

\maketitle

\begin{abstract}
\input{sections/abstract.tex}
\end{abstract}


\section{Introduction}
\subsection{Motivation}
\input{sections/intro/motivation.tex}
\subsection{Related Work and Main Contributions}
\input{sections/intro/contributions.tex}
\subsection{Notation}
\input{sections/intro/notation.tex}

\section{Preliminaries} \label{sec:prelimiaries}
\subsection{System Model}\label{sec:sys_model}
\input{sections/preliminaries/sys_model.tex}
\subsection{Metrics}
\input{sections/preliminaries/metrics.tex}
\section{Jammer Model}
\input{sections/preliminaries/jammer_model.tex}

\section{Resilient Receiver Design} \label{sec:transceiver}
\subsection{General Considerations}\label{sec:look_sinr}
\input{sections/resilient_design/look_sinr.tex}
\input{sections/resilient_design/separate_design.tex}


\section{Numerical Results} \label{sec:results}
\subsection{Methodology}
\input{sections/results/methodology.tex}
\subsection{Simulation Studies}
\input{sections/results/simulation.tex}

\section{Conclusions} \label{sec:conclusions}
\input{sections/conclusions.tex}
\bibliography{IEEEabrv, references}
\bibliographystyle{IEEEtran}
\end{document}

%% file: sections/preamble.tex
\DeclareMathOperator*{\trace}{tr}
\newcommand{\tr}[1]{\trace\left(#1\right)}

\newcommand{\normsq}[1]{\left\lVert#1\right\Vert_{2}^{2}}
\newcommand{\frob}[1]{\left\lVert#1\right\Vert_{F}^{2}}

\newcommand{\cov}[1]{\bm{C}_{#1}}

\newcommand{\gauss}[1]{\mathcal{N}\left(\bm{0}, #1\right)}

\newcommand{\eye}{\bm{\mathrm{I}}}

\newcommand{\herm}{^{\mathsf{H}}}
\newcommand{\tran}{^{\mathsf{T}}}

\newcommand{\C}{\mathbb{C}}
\newcommand{\R}{\mathbb{R}}
\newcommand{\ex}{\mathrm{e}}
\newcommand{\jma}{\mathrm{j}}

\newcommand{\abg}{\bm{A}_B(\bm{\theta}_G)}

\newcommand{\x}{\bm{x}}
\newcommand{\y}{\bm{y}}
\newcommand{\ch}{\bm{H}}
\newcommand{\chk}{\ch_{k}}
\newcommand{\cg}{\bm{G}}
\newcommand{\z}{\bm{z}}

\newcommand{\nak}{N_{A_{k}}}

\newcommand{\wa}{\bm{w}}
\newcommand{\wak}{\wa_{k}}
\newcommand{\xk}{\x_{k}}

\newcommand{\shb}{\hat{s}}
\newcommand{\shbk}{\shb_{k}}
\newcommand{\vb}{\bm{v}}
\newcommand{\vbk}{\vb_{k}}

\newcommand{\vvbk}{\bm{V}_{k}}

\newcommand{\kpr}{k^{\prime}}
\newcommand{\chkp}{\ch_{\kpr}}
\newcommand{\wakp}{\wa_{\kpr}}
\newcommand{\interf}{\sum\limits_{\kpr\neq k}
\chkp\wakp\wakp\herm\chkp\herm}

\newcommand{\useful}{\chk\wak\wak\herm\chk\herm}

\newcommand{\ar}{\bm{a}_{B}}

\newcommand{\atj}{\bm{a}_{J}}

\newcommand{\br}[1]{\left( #1 \right)}

\newcommand{\brcur}[1]{\left\{ #1 \right\}}

%% file: sections/abstract.tex
We address the resilience of future 6G MIMO communications
by considering an uplink scenario where multiple legitimate transmitters try to communicate with a base station in the presence of an adversarial jammer.
The jammer possesses full knowledge about the system and the physical parameters of the legitimate link, 
while the base station only knows the UL-channels and the angle-of-arrival (AoA) of the jamming signals.
Furthermore, the legitimate transmitters are oblivious to the fact that jamming takes place, thus 
the burden of guaranteeing resilience falls on the receiver.
For this case we derive one optimal jamming strategy that aims to minimize the rate of the strongest user
and multiple receive strategies, 
one based on a lower bound on the achievable signal-to-interference-to-noise-ratio (SINR),
one based on a zero-forcing (ZF) design, and one based on a minimum SINR constraint.
Numerical studies show that the proposed anti-jamming approaches 
ensure that the sum rate of the system is much higher than without protection, 
even when the jammer has considerably more transmit power and even if the jamming signals
come from the same direction as those of the legitimate users.  

%% file: sections/intro/motivation.tex
While the first generations of mobile communications standards concerned themselves with increasing data-rates
and connectivity, the latter generations focused on a wide array of use-cases and applications.
With the currently deployed 5G standard, very high data rates, reliability, low latency and massive connectivity
between a very large number of entities are made possible, enabling applications such as the Tactile Internet
for businesses, virtual reality and advanced, remote-controlled robotics \cite{tactile_internet}.
In the future 6G standard, data rates and time sensitivity orders of magnitude higher than what are currently available, 
as well as connectivity for all things and digital twins are envisioned \cite{6g_overview}. 

One of the key drivers towards enabling this massive digital transformation is integrated, or joint communication and 
sensing (JCAS)\cite{jcas_zhang}, which offers radio sensing services
along with communications capabilities, all under a unified hardware platform \cite{zhang2021overview}.
With the emergence of the virtualization not only of data, but of whole physical objects at once,
the question of trustworthiness becomes crucial \cite{fettweis_boche_resilience2022}, 
which comprises, among other things the resilience of a system.

While jamming can also be friendly \cite{friendly_jamming_usrp}, 
a malicious jammer, who has access to all physical and system parameters of the environment,
such as communication protocols, signal processing, physical channels, etc. is considered. 
This is also known as \textit{worst-case} or \textit{smart jammer}.
The legitimate entities on the other hand only have 
access to their own system and physical parameters, 
and have very limited knowledge about the adversarial entity.
In \cite{fettweis_boche_resilience2022} it was shown that in the classical communication setting (without sensing),
the question whether a denial-of-service attack has taken place is undecidable on a Turing machine \cite{boche2020}. 
This means that the coding layer can not ensure resilience against such attacks at the resource allocation level, 
and that other coordination mechanisms, also known as common randomness (CR), are required \cite{boche2020}. 
In this paper, we explore how to embed sensing information into a system architecture, which needs to be resilient by design.
Our main goal is to optimally leverage sensing information and the spatial structure of the radio channel 
in order to design transmit and receive filters which are resistant against worst-case jammers.
Note that this approach might naturally be interpreted in the CR context, since the AoAs assure the coordination resources.   

%% file: sections/intro/contributions.tex
While anti-jamming is a widely studied subject, to our knowledge there are no works which address the problem of 
designing resilient transceivers against a jammer who employs his optimal strategy by leveraging 
sensing information and the spatial structure of the radio channel.
In \cite{anti_jamming_planar} the authors propose a database approach to nullifying signals coming from the jammer
direction using massive planar arrays, and by only considering the jammer power and not the its transmit strategy, 
In \cite{rl_jamming_noma} and \cite{rl_jamming_irs}, 
the authors use reinforcement-learning (RL) to optimize various system parameters under the simplifying assumption
that the jammer transmits uniformly over all streams. 

The characterization of a worst-case jammer in the context of MIMO communications has been done 
in terms of the optimal jammer strategy, albeit without considering the spatial structure of the radio channel.
The authors in \cite{corr_jamming} derive the optimal transmit and jammer strategies
using knowledge of the transmitted symbols strategy at the jammer and show that under certain conditions, this does not affect jammer performance. 
In \cite{worst_case_jamming_mimo} the authors prove that there exists a lower bound on the achievable rate of communication 
between the legitimate party, which does not depend on the jammer setup. 
Furthermore, they characterize and give a closed form solution of the optimal jamming strategy in the so-called 
\textit{jammer-dominant regime} 
while in \cite{disrupting_mimo_comms} the authors derive the optimal jamming strategy in closed form, which minimizes
the signal-to-interference-and-noise-ratio (SINR) at the legitimate receiver in a MIMO system.
At last, in \cite{mimo_jammer_mac_bc} the authors use game theoretical tools to approach
the problem of optimal resource allocation in the MIMO multiple access (MAC) and broadcast channels (BC) under jammer conditions,
and show that the Nash equilibria of the resulting games always exist.

The only link between the concept of jamming and joint communication and sensing (JCAS) we are aware of, are the works of 
\cite{su_pls_bf} and \cite{pls_jcas_su}, where the authors employ knowledge about an eavesdroppers location 
in order to send artificial noise in its direction, therefore increasing the secrecy rate.
Thus, our contributions are as follows:
\begin{itemize} 
    \item We construct a communication model which exploits the spatial structure of the radio channel and sensing information for a multi-user MIMO MAC scenario.
    \item We propose an optimal jamming strategy which minimizes the achievable rate of the strongest user, thus being robust against
    Successive Interference Cancellation (SIC) \cite{sic}.
    \item Next we propose anti-jamming receive filters at the BS which only use AoA information. 
    The proposed designs are based on a lower bound on the individual per-stream rate, a zero-forcing, and a minimum SINR constraint respectively.
    \item We demonstrate by numerical experiments, that we can ensure constant, satisfactory performance regardless of the number of antennas at the jammer.
    We also demonstrate numerically that the achieved sum-rate mainly depends on the AoA of the jamming signals and on the jamming power,
    and that the proposed designs can ensure a non-zero sum rate even in the most extreme cases.
\end{itemize}
 

%% file: sections/intro/notation.tex
Throughout this paper we denote the sets of natural, real and complex numbers by $\mathbb{N}$, $\R$ and $\C$, respectively. 
We use lower case letters for scalars $x$, bold lower-case letters for vectors $\x$ and
bold upper-case letters for matrices $\bm{X}$. We write sets as $\{x_{i}\}_{i=1}^{S}$, where $i \in \mathbb{N}$ is the index
and $S$ is the cardinality of the set. 
The hermitian transponse and inverse of a matrix $\bm{X}$ are denoted by $\bm{X}\tran$, $\bm{X}\herm$,
and $\bm{X}^{-1}$, respectively. 
The trace, Frobenius norm and rank of a matrix $\bm{X}$ are written as $\tr{\bm{X}}$, $\lVert \bm{X} \rVert_{F}$
and $\text{rank}(\bm{X})$. 
The symbol $\succcurlyeq$ denotes the semi-ordering relationship on the cone of positive semi-definite matrices.
We use $\lVert \x \rVert_{2}$ to denote the Euclidean norm of a vector
and $\bm{e_{k}}$ for the $k$-th canonical basis vector in $\C^{N}$. 
Here, $\eye$ denotes the identity matrix.
By $\x \sim \gauss{\cov{\x}}$ we say the random vector $\x$ is a zero-mean, proper Gaussian random variable
with covariance matrix $\cov{\x}$, and by $\mathcal{U}(a, b)$ we denote the uniform distribution on the interval $[a, b]\in\R$.
Lastly, the mutual information between the two random variables $\x$ and $\y$ is denoted as $I(\x; \y)$ 
and the real part of $\z\in\C^{N}$ as $\Re\brcur{\z}$.

%% file: sections/preliminaries/sys_model.tex
We assume a setup consisting of $K$ legitimate transmitters, a legitimate receiver / base station (Bob) and jammer (Jimmy).
The transmitters send their data streams $s_{k} \sim \mathcal{N}(0, 1)$ by applying precoding with $\wak\in\C^{\nak}$ through $\nak$ antennas.
The precoding vectors $\wak$ satisfy the power constraint $\normsq{\wak} \leq P_{A_{k}}$.
The resulting signals $\xk = \wak s_{k}$ propagate through the legitimate channels $\chk\in\mathbb{C}^{N_{B}\times \nak}$ and arrive at Bob's $N_{B}$ antennas,
where we assume $N_{B} \geq K + 1$. 
The signal is corrupted by white noise $\bm{n}\sim\gauss{\sigma^2\eye}$, as well as by the jammer signal $\z \sim \gauss{\cov{\z}} \in\C^{N_{J}}$, which propagates through the channel $\cg\in\C^{N_{B}\times N_{J}}$,
and satisfies the power constraint $\tr{\cov{\z}} \leq P_{J}$.
At last, Bob applies the equalizers $\vbk\herm \in \C^{1 \times N_{B}}$ to build an estimate $\shbk$ for each the transmitted streams $s_{k}$. 
More formally:
\begin{align}
    &\xk = \wak s_{k} \\
    &\y = \sum_{k=1}^{K}\chk\xk + \cg\z + \bm{n} \\
    &\shbk = \vbk\herm\y = \vbk\herm\left(\sum_{k=1}^{K}\chk\wak s_{k}\right) + \vbk\herm\cg\z +  \vbk\herm\bm{n}
\end{align}
The legitimate channels $\chk$ and the jammer channel $\cg$ are modelled as spatial beam-space channels, namely
\begin{align}
    \label{eq:beamspace}
    \chk &= \sum_{l=1}^{L_{H_{k}}} b_{H_{k}, l} \bm{a}_{B}(\theta_{H_{k},l}) \bm{a}_{A_{k}}(\psi_{H_{k},l})\herm \\
    \cg &= \sum_{l=1}^{L_{G}} b_{G, l} \bm{a}_{B}(\theta_{G, l}) \bm{a}_{J}(\psi_{G,l})\herm ,
\end{align}
with $L_{H_{k}}$, $L_{G}$ being the number of resolvable paths for each channel, $\bm{a}_{S},\, S\in\{A, B, J\}$ the steering vectors,
and $\theta_{\cdot, l}$, $\psi_{\cdot, l}$, $b_{\cdot, l}$ the angles of arrival, angles of departure and path gain,
corresponding to the $l$-th resolvable path in each channel, respectively.

For simplicity of analysis, we assume a 2D geometry, and thus all parties employ uniform linear arrays (ULA), with the steering vector given by 
\begin{align}
    \bm{a}_{S}(\theta) = \begin{bmatrix}
        1&
        \ex^{-\jma \frac{2\pi d}{\lambda_{c} } \cdot \sin\theta}&
        \dots&
        \ex^{-\jma \frac{2\pi d}{\lambda_{c} } \cdot (N-1) \sin\theta}
    \end{bmatrix} \tran 
\end{align}
where $\lambda_{c}$, $d$ and $N$ denote the wavelength, element spacing, and the number of elements.
Defining the matrices
\begin{align}
    \label{eq:ula}
        \bm{A}_B(\bm{\theta}_{G}) &= \begin{bmatrix}
        \ar(\theta_{G,1}) & \dots & \ar(\theta_{G,L_{G}})
        \end{bmatrix}\\
        \bm{A}_J(\bm{\psi}_{G}) &= \begin{bmatrix}
        \atj(\psi_{G,1}) & \dots & \atj(\psi_{G,L_{G}})
        \end{bmatrix}\\
        \bm{B}_{G} &= \text{diag}\{b_{G, l}\}_{l=1}^{L_{G}}  
\end{align}
with $\bm{A}_B(\bm{\theta}_{G}) \in \C^{N_{B} \times L_{G}}$, $\bm{A}_J(\bm{\psi}_{G}) \in \C^{N_J \times L_{G}}$, 
$\bm{B}_{G} \in \C^{L_{G}\times L_{G}}$, the jammer channel $\cg$ becomes
\begin{equation}
    \label{eq:jammer_channel}
    \cg = \bm{A}_B(\bm{\theta}_G)\bm{B}_{G}\bm{A}_J(\bm{\psi}_{G})\herm.
\end{equation}

%% file: sections/preliminaries/metrics.tex
The achievable rates $R_{k}^{A}$ for each of the $K$ users under \textit{no particular decoding order}
at the BS and the per-stream achievable rates $R_{B}^{k}$ are given by
\begin{align}
    R_{k}^{A} &= I(\y; s_{k}) = \log(1 + \gamma_{k}^{A}) \\
    R_{k}^{B} &= I(\shbk; s_{k}) = \log(1 + \gamma_{k}^{B}),
\end{align}
with $\gamma_{k}^{A}$, $\gamma_{k}^{B}$ as in equations \ref{eq:sinra} and \ref{eq:sinrb}.
\begin{figure*}[t]
    \begin{align} \label{eq:sinra}
    \gamma_{k}^{A} &= \wak\herm  \chk\herm 
    \br{\interf + \cg\cov{\z}\cg\herm + \sigma^2\eye}^{-1} 
    \chk \wak \\
    \label{eq:sinrb}
    \gamma_{k}^{B} &= \dfrac{
        \vbk\herm \useful \vbk
    }{\vbk\herm \br{\interf + \cg\cov{\z}\cg\herm + \sigma^2\eye} \vbk} 
    \end{align}
    \hrulefill
\end{figure*}
The achievable sum-rates $R^{A}$, $R^{B}$ are then given by 
summing over all $R_{k}^{A}$'s and $R_{k}^{B}$'s respectively.

%% file: sections/preliminaries/jammer_model.tex
We now present the smart jammer model used throughout the paper.
We assume the jammer has access to all system and physical parameters of all links, 
i.e. $\{\wak, \chk\, \vbk\}_{k=1}^{K}$, $\cg$, $\sigma^{2}$, but not to the transmitted symbols $s_{k}$, since it has been shown in \cite{corr_jamming}
that knowledge about $s$ does not increase jammer performance if $\{\wak, \chk\}_{k=1}^{K}$ are known.
The next assumption is that the jammer has more antennas than all parties, 
i.e. $N_{J} \geq N_{A_{k}}, N_{B} \, \forall k$, as well as more transmit power than all transmitters $P_{J} \geq P_{A_{k}} \, \forall k$.
Finally, we assume that the disturbing transmission takes place in the so-called \textit{jammer-dominant regime}.
This concept was rigorously formalized in \cite{worst_case_jamming_mimo}, and roughly speaking, means that
the jammer power is much higher than the noise power, i.e. $P_{J} \gg \sigma^2$.

The goal of the jammer is to minimize the sum-rate $R^{A}$ of all transmitters under the power constraint $P_{J}$.
In order to achieve this, we consider the inequality
\begin{equation}
    R^{A} = \sum_{k=1}^{K} R_{k}^{A} \leq K\max_{k}R_{k}^{A} .
\end{equation}
Thus, if the rate of the ``strongest'' user can be brought to $0$, then the sum rate is also $0$.
Since $\log$ is a monotonically increasing function, the problem can be cast as
\begin{align}
    \label{eq:opt_jammer}
        \cov{\z}^{*} = \arg\min_{
        \substack{\cov{\z} \succcurlyeq \bm{0},\, \cov{\z} = \cov{\z}\herm \\
        \tr{\cov{\z}} \leq P_{J}}
        } \max_{k} \gamma_{k}^{A} .
\end{align}
This is a convex optimization problem, since both the objective function 
\footnote{The maximum over a family of convex functions is itself convex, the function $\x\herm\bm{B}^{-1}\x$ is convex in symmetric positive semidefinite $\bm{B}$ \cite{boyd_cvx}.} 
and the constraints are convex. 
Note that we do not have to consider the equalizer at the receiver, since $R_{k}^{A} \geq R_{k}^{B} \, \forall k$ by standard information theoretical arguments.
Furthermore, note that this jamming strategy can be greatly simplified if Successive Interference Cancellation (SIC) \cite{sic} is used at the receiver, 
since in this case, only the rate of one user needs to be minimized, rendering the problem equivalent to that in \cite{disrupting_mimo_comms}.
In the following sections, we will see how to ensure protection 
over a wide range of channel conditions when dealing with a jammer employing the strategy presented in  \eqref{eq:opt_jammer}.

%% file: sections/resilient_design/look_sinr.tex
We assume the legitimate receiver has perfect knowledge of the channels and signal processing $\{\wak, \chk\}_{k=1}^{K}$ at the transmitters,   
and only has access to the AoAs $\bm{\theta}_{G} = \{\theta_{G, l}\}_{l=1}^{L_{G}}$ of the impinging jammer signals.
Furthermore, the legitimate transmitters do not know the communication is jammed,
and do not cooperate with each other. 
That being said, we assume for simplicity of analysis that the transmitters employ singular value decomposition (SVD),
namely
\begin{equation}
    \wak = \mathcal{V}_{\text{max}}(\chk),
\end{equation}
where $\mathcal{V}_{\text{max}}(\chk)$ denotes the right-singular vector corresponding to the 
largest singular value of the matrix $\chk$.

In the following, we will concentrate on the receiver and derive a bound on the individual per-stream rate $R_{k}^{B}$, 
which depends, up to a constant, 
solely on its setup. We will then use this lower bound in our designs.

Let $\vvbk = \vbk\vbk\herm$ and consider 
\begin{align} \nonumber
    \vbk\herm \left( \cg\cov{\z}\cg\herm + \sigma^2\eye \right)\vbk &= \tr{\left( \cg\cov{\z}\cg\herm + \sigma^2\eye \right) \vvbk}\\
    \label{eq:denom_trace}
    &= \tr{\cg\cov{\z}\cg\herm \vvbk + \sigma^2\vvbk }.
\end{align}
Plugging in the expression in \eqref{eq:jammer_channel} and 
introducing the matrix $\bm{E} = \bm{B}_{G}\bm{A}_J(\bm{\psi}_{G})\herm\cov{\z}^{1/2}$, we obtain
for the first term inside the trace
\begin{align}
    &\tr{\cg\cov{\z}\cg\herm \vvbk } = \tr{\abg \bm{E}\bm{E}\herm \abg\herm \vvbk} \\
    \label{eq:step1}
    &= \tr{\bm{E}\bm{E}\herm \abg\herm \vvbk \abg} \\
    \label{eq:step2}
    &\leq \frob{\bm{E}} \tr{\abg\herm \vvbk \abg} \\
    \label{eq:step3}
    &\leq \eta \tr{\abg\herm \vvbk \abg},
\end{align}
with $\eta = P_{J}N_{J}L_{G}\frob{\bm{B}_{G}}$.
\eqref{eq:step1} follows from the commutativity of trace, 
\eqref{eq:step2} follows from the fact that $\tr{\bm{AB}} \leq \tr{\bm{A}}\tr{\bm{B}}$ 
for $\bm{A}$, $\bm{B}$ symmetric positive-semidefinite \cite[Section 1]{trace_ineq}, and lastly,
\eqref{eq:step3} follows from applying the sub-multiplicativity of the Frobenius norm twice, 
and from equations \eqref{eq:beamspace} and \eqref{eq:ula}, respectively.
With the shorthands,
\begin{align}
    \label{eq:sh1}
    \bm{A}_{k} &= \useful \\ \label{eq:sh2}
    \bm{B}_{k} &= \interf + \sigma^2 \eye,
\end{align}
we have
\begin{align} \label{eq:lb}
    \gamma_{k}^{B} \geq 
    \dfrac{
        \vbk\bm{A}_{k}\vbk
    }{\vbk\herm \br{\bm{B}_{k} + \eta\abg\abg\herm} \vbk}
    \overset{\Delta}{=} \tilde{\gamma}_{k}^{B} .
\end{align}
We note that equality in \eqref{eq:lb} is achieved if $P_{J} = 0$ (no jammer) or if $\bm{A}_B(\bm{\theta}_G)\herm \vvbk = \bm{0}$, i.e.
equalizer $\vbk$ lies in a subspace orthogonal to the array manifold $\bm{A}_B(\bm{\theta}_G)$.

We note that aside from $\eta$, all other terms in \eqref{eq:lb} are known at the receiver.
In the jammer-dominant regime the term $\eta \tr{\abg\herm \vvbk \abg}$ is much greater
that $\sigma^2$ (since it scales linearly in $P_{J}$ and $N_{J}$), effectively dominating the influence of the noise.


%% file: sections/resilient_design/separate_design.tex
\subsection{Closed-Form Design}\label{subsec:closed_form}
The receivers' goal is to ensure reliable communication for all uplink participants,
i.e. we are interested in maximizing the individual per-stream SINRs $\gamma_{k}^{B}$ or 
at least ensure a certain quality of service (QoS), e.g. the $\gamma_{k}^{B}$'s lie above a given threshold $\gamma_{0}$.

The maximization of the lower bound in Eq. \eqref{eq:lb} is a Rayleigh quotient maximization problem with the
standard solution being given by:
\begin{align}
    \label{eq:maxsinr}
    \vbk = \br{\bm{B}_{k} + \eta\abg\abg\herm}^{-1}\chk\wak
\end{align} 
for all $k = 1, \dots, K$. 
This is the most simple way to leverage AoA information at the receiver, since the solution is given in closed form
and computationally tractable, since $\abg\abg\herm$ can be easily computed without matrix multiplications 
by only considering the phase shifts between the impinging jamming signals.

Note, that the resulting filter still depends on $\eta$, which in general is not available at the receiver.
The most easy way to address this problem is to consider $\eta$ as a hyperparameter which controls how much the
jamming signals are suppressed. 
Indeed, this is justified by the fact that in the jammer-dominant regime, 
the contribution of the noise variance $\sigma^{2}$ is small compared to the one of the jammer, i.e
$\vbk\herm \br{\eta\abg\abg\herm + \sigma^{2}\eye} \vbk \approx \eta\vbk\herm \abg\abg\herm \vbk$.
In the following, we will propose other methods to overcome this dependency.

\subsection{Zero Forcing Design}
As it can be seen in Section \ref{subsec:closed_form}, the main contribution to the jamming signals is given
by the term 
\begin{align}
    \beta_{k} = \normsq{\bm{A}_B(\bm{\theta}_G)\herm \vbk} 
\end{align}
which can be naturally interpreted as the receive beampattern of the receiver antenna array \cite{van2004optimum}. 
This indicates how much the signals coming from the jammer direction are emphasized, thus 
the anti-jamming criterion can be formalized as a beampattern minimization problem in the same spirit as
\cite{mu_mimo_radar_com,su_pls_bf,jcas_sp_survey}.
Furthermore, a natural way to ensure good per-stream SINRs for all users is to remove the interference,
which can be realized with a zero-forcing (ZF) design. 
Thus, the beampattern minimization problem under ZF constraint can be formulated analogously to \cite{mu_mimo_radar_com} as
\begin{align}
    \label{eq:bpmin_zf}
    \min_{\substack{
        \vbk\\
        k = 1, \dots, K}
    }\sum_{k=1}^{K}\beta_{k} \quad \text{s.t.} \quad \bm{P}\herm \vbk = \bm{e}_{k} \, \forall k
\end{align}
where we defined 
\begin{equation}
    \bm{P} = \begin{bmatrix}
        \ch_{1}\wa_{1} & \dots & \ch_{K}\wa_{K}
    \end{bmatrix} \in \C^{N_{B} \times K}
\end{equation}
and $\bm{e}_{k}$ is $k$-th standard basis vector in $\R^{K}$.
Note that this is a convex optimization problem easily solvable in polynomial time. 

Furthermore, if the matrix $\abg\abg\herm$ has full-rank, we can also give an analytic solution.
Letting $\bm{X} = \abg\abg\herm$, we construct the Lagrangian
\begin{align}
    \mathcal{L}\left(\vbk, \bm{\mu}_{k}\right) = 
    \sum_{k=1}^{K} \vbk\herm\bm{X}\vbk + 2\Re\brcur{\bm{\mu}_{k}\herm \br{\bm{P}\herm \vbk - \bm{e}_{k}}}
\end{align}
where $\bm{\mu}_{k}\in\C^{K}$ are the Lagrange multipliers.
Since the ZF-constraint is an equality constraint, the Karush-Kuhn-Tucker conditions are both necessary and sufficient.
Computing the gradients wrt. $\vbk$ and $\bm{\mu}_{k}$, setting them to $\bm{0}$, and solving for $\vbk$,
we obtain
\begin{align}
    \label{eq:bpmin_zf_analytic}
    \vbk = \bm{X}^{-1} \bm{P} \br{\bm{P}\herm \bm{X}^{-1} \bm{P}}^{-1} \bm{e}_{k}
\end{align}
By standard linear algebra $\bm{X}$ has full rank iff $L_{G}\geq N_{B}$, which might pose a problem if the propagation channel
of the jammer shows $L_{G} < N_{B}$ dominant paths. 
In that case, one can easily ``pad'' the resolved AoAs $\bm{\theta_{G}}$ with 
a set of different angles $\{\phi_{G, l}\}_{l=1}^{N_{B}-L_{G}}$ thus enlarging the grid over which the beampattern should be minimized.
This idea can be naturally extended to the designs considered in Sections \ref{subsec:closed_form} and \ref{subsec:min_sinr}.
\subsection{Minimum SINR Design}\label{subsec:min_sinr}
Another way to approach this problem is to minimize the influence of the jamming signals, 
while ensuring a minimum SINR $\gamma_{0}$ in the jammer-free case.
Using this method one can address the disadvantages ZF-filters generally have in the low SNR regime, i.e. $\sigma^{2} \gg 0$.
Rewriting the beampattern $\beta_{k}$ as
\begin{align}
    \beta_{k} = \normsq{\bm{A}_B(\bm{\theta}_G)\herm \vbk} = \tr{\bm{A}_B(\bm{\theta}_G)\herm \vvbk \bm{A}_B(\bm{\theta}_G)}
\end{align}
we cast the resulting problem as
\begin{align}
    \label{eq:bpmin}
    \min_{
        \substack{\vvbk \succcurlyeq \bm{0}\, \vvbk = \vvbk\herm \\ \text{rank}(\vvbk) = 1 \\
        k = 1, \dots, K}
    } \,
    \sum_{k=1}^{K}\beta_{k} \quad \text{s.t.} \quad \dfrac{\tr{\bm{A}_{k}\vvbk}}{\tr{\bm{B}_{k}\vvbk}} \geq \gamma_{0}
\end{align}
with $\vvbk = \vbk \vbk \herm$ as before and $\bm{A_{k}}$ and $\bm{B_{k}}$ as in equations \eqref{eq:sh1} and \eqref{eq:sh2}.
Note that this problem is non-convex due to the rank-$1$ constraint on $\vvbk$. 
In order to efficiently solve this problem, we apply semidefinite relaxation \cite{sdr}, i.e. 
we relax the rank-$1$ constraint, 
solve the resulting convex problem and recover the solution via eigenvalue decomposition (EVD).

%% file: sections/results/methodology.tex
In this section we present the performance of the proposed designs with respect to the jammer setup and channel conditions.
To this end, we consider three parameters, namely the number of antennas at the jammer $N_J$,
the power budget of the jammer $P_J$, and the AoAs $\bm{\theta}_{G}$ of the impinging signals.
We shall compare the sum-rates obtained by the proposed designs with the sum-rates in the jammer-free case,
as well as in the case when the parties do not ensure any jammer protection.
We consider a setup of $K=3$ users, each with $\nak = 8$ and $P_{A_{k}} = 5\si{\dBm}$.
The number of antennas at the legitimate transmitter $N_{B}$ is kept fixed at $16$, 
and the noise variance $\sigma^{2} = -10 \si{\decibel}$.
In order to generate the channel matrices, we draw all path gains and AoDs in \eqref{eq:beamspace}
randomly from $\mathcal{N}(0, 1)$ and $\mathcal{U}(-5^\circ, 5^\circ)$.
The AoAs are generated as 
$\theta_{G, l} = \theta_{J} + \phi_{l}$ and $\theta_{H_{k}, l} = \theta_{A_{k}} + \omega_{l}$,
with $\phi_{l}$ and $\omega_{l}$ drawn i.i.d from $\mathcal{U}(-5^\circ, 5^\circ)$, and $\theta_{A_{k}} = \{-10^{\circ}, 0^{\circ}, 10^{\circ}\}$ from each other.
The direction dependency of the channels are then characterized by the central AoA
$\theta_{J}$ of the impinging jamming signal.

%% file: sections/results/simulation.tex
\begin{figure}
    \input{sections/figures/sum_rates_avg_ant.tex}
    \centering
    \caption{Averaged sum-rates for different number of antennas at the jammer, 
    $\gamma_{0} = 20\si{\decibel}$, $\eta = 1$.}
    \label{fig:antennnas}
\end{figure}
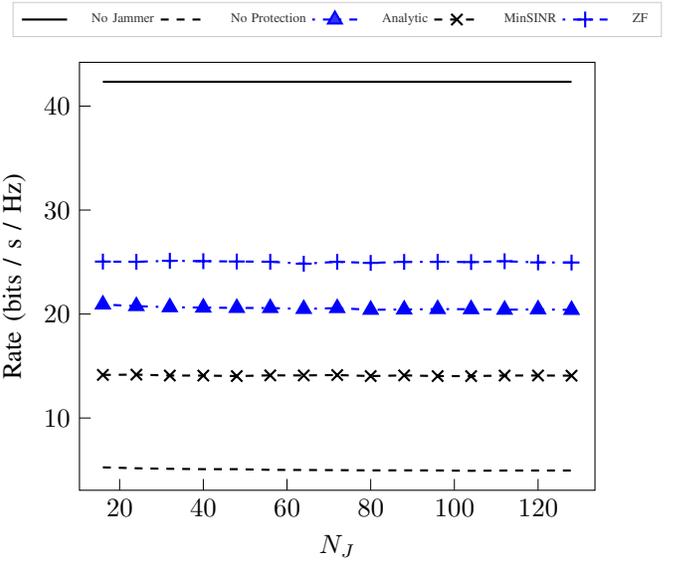
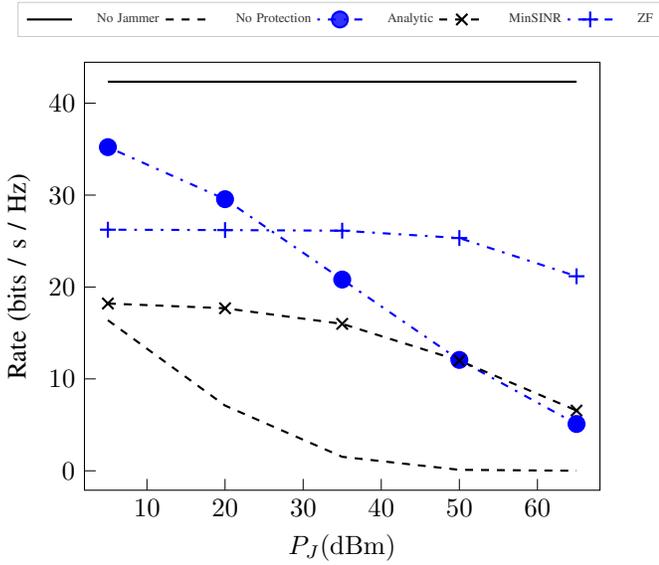
\begin{figure}
 
    \input{sections/figures/sum_rates_avg_pow.tex}
    \centering
    \caption{Averaged sum-rates for different jammer powers,
    $\gamma_{0} = 20\si{\decibel}$, $\eta = 1$.}
    \label{fig:powers}
\end{figure}
\begin{figure}
    \input{sections/figures/sum_rates_avg_aoa.tex}
    \centering
    \caption{Averaged sum-rates for different directions of arrival, 
    $\gamma_{0} = 20\si{\decibel}$, $\eta = 1$.}
    \label{fig:aoas}
\end{figure}
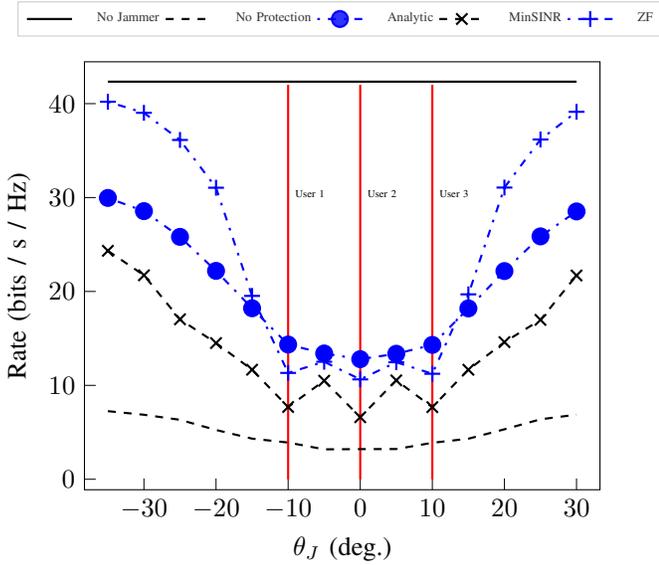

We first start by assessing the influence of the number of antennas at the jammer on the communication quality.
To this end, we vary the number of antennas at the jammer $N_{J}$ and average the computed sum-rates over $\theta_{J}$ and $P_{J}$.
We first observe that the obtained sum rates are independent of the number of antennas at the jammer,
since the performance curves remain relatively flat along the whole range.
All of the proposed designs offer a much higher rate than in the cases without protection, 
with a minimum gain of $10 \text{bits}/\si{\second}/\si{\hertz}$.
In terms of performance ranking, the ZF-based design from equation \eqref{eq:bpmin_zf} (denoted as ``ZF'' in Figures \ref{fig:antennnas}-\ref{fig:aoas}) performs best, 
followed by the optimal analytic filter based on the surrogate objective in equation \eqref{eq:maxsinr} (``Analytic'' in Figures \ref{fig:antennnas}-\ref{fig:aoas}). 
The filter based on the minimum QoS (Equation \eqref{eq:bpmin}), denoted ``MinSINR'' in the plots, comes last. 

We attribute this ranking to the fact that the filters ``Analytic'' and ``MinSINR'' are much more sensitive to the jamming power, 
since the former is coupled to it by the constant $\eta$, and the latter is based on an objective, 
that should ensure a minimum SINR at all times. 
This might not be fully-possible when jamming powers are high.
In order to test this conjecture, we vary the jammer power $P_{J}$ and average the computed sum-rates over $\theta_{J}$ and $N_{J}$.
Indeed, while the filters based on the minimum SINR criterion and analytic formula decline sharply over the studied range,
the filter based on the ZF criterion stays approximately constant over the whole range, only declining for high jammer powers.

Finally, we plot the performance of the proposed designs with respects to the AoA $\theta_{J}$ of the impinging jamming signals in Figure \ref{fig:aoas}.
Note, that the AoAs of the users are marked with a vertical, red line. 
We observe that the sum-rate decreases as the AoAs of the jamming signals become similar to those of the legitimate parties.
We furthermore observe, that the jammer can not drive the sum-rate to $0$ in the studied configuration.
In terms of performance, the filter based on the analytic formula shows slightly better performance, than the ZF approach, 
in the case that the jamming and legitimate AoAs are similar. 

%% file: sections/figures/sum_rates_avg_ant.tex
\begin{tikzpicture}

\definecolor{darkgray176}{RGB}{176,176,176}
\definecolor{lightgray204}{RGB}{204,204,204}

\begin{axis}[
legend cell align={left},
legend style={
  fill opacity=0.8,
  draw opacity=1,
  text opacity=1,
  at={(0.5, 1.1)},
  anchor=center,
  draw=lightgray204,
  font=\tiny,
},
legend columns=-1
tick align=outside,
tick pos=left,
x grid style={darkgray176},
xlabel={\(\displaystyle N_{J}\)},
xmin=10.4, xmax=133.6,
xtick style={color=black},
y grid style={darkgray176},
ylabel={Rate (bits / s / Hz)},
ymin=3.0497300477319, ymax=44.2108647753323,
ytick style={color=black}
]
\addplot [thick, black]
table {%
16 42.3399041058959
24 42.3399041058959
32 42.3399041058959
40 42.3399041058959
48 42.3399041058959
56 42.3399041058959
64 42.3399041058959
72 42.3399041058959
80 42.3399041058959
88 42.3399041058959
96 42.3399041058959
104 42.3399041058959
112 42.3399041058959
120 42.3399041058959
128 42.3399041058959
};
\addlegendentry{No Jammer}
\addplot [thick, black, dashed]
table {%
16 5.2536947163718
24 5.16508222916462
32 5.11311332665452
40 5.07923113597069
48 5.07942518578309
56 5.01339945529517
64 4.99929276180926
72 4.98551977726919
80 4.96412839401032
88 4.96709109732642
96 4.95831453303444
104 4.92069071716828
112 4.94945298779099
120 4.95084775535114
128 4.95342299600632
};
\addlegendentry{No Protection}
\addplot [thick, blue, dash pattern=on 1pt off 3pt on 3pt off 3pt, mark=triangle*, mark size=3, mark options={solid}]
table {%
16 20.9187031298759
24 20.762563375725
32 20.6551604538096
40 20.619596124442
48 20.5864629029313
56 20.5787236614299
64 20.4893331024003
72 20.5718298865597
80 20.4088233515677
88 20.4381966681961
96 20.4751978689041
104 20.4644448582667
112 20.4140817530421
120 20.4513178159494
128 20.397562981978
};
\addlegendentry{Analytic}
\addplot [thick, black, dashed, mark=x, mark size=3, mark options={solid}]
table {%
16 14.145307701722
24 14.1726338750413
32 14.0851249537213
40 14.0809736509022
48 14.0342045841602
56 14.101573653538
64 14.0965890830918
72 14.1326333951599
80 14.0335783672618
88 14.0993986491415
96 14.031826838323
104 14.0317147850664
112 14.0839231807054
120 14.0882853080797
128 14.0727234684127
};
\addlegendentry{MinSINR}
\addplot [thick, blue, dash pattern=on 1pt off 3pt on 3pt off 3pt, mark=+, mark size=3, mark options={solid,rotate=270}]
table {%
16 25.0449086660649
24 25.0256651749056
32 25.1200250532137
40 25.0836655624462
48 25.0537873658047
56 25.0294907019906
64 24.8421366738044
72 25.0190949559919
80 24.9198596472099
88 25.0160739983922
96 25.0232735507865
104 24.9987709224178
112 25.0784790644852
120 24.9672458879809
128 24.9538136396102
};
\addlegendentry{ZF}
\end{axis}

\end{tikzpicture}

%% file: sections/figures/sum_rates_avg_pow.tex
\begin{tikzpicture}

\definecolor{darkgray176}{RGB}{176,176,176}
\definecolor{lightgray204}{RGB}{204,204,204}

\begin{axis}[
legend cell align={left},
legend style={
  fill opacity=0.8,
  draw opacity=1,
  text opacity=1,
  at={(0.5, 1.1)},
  anchor=center,
  draw=lightgray204,
  font=\tiny,
},
legend columns=-1
tick align=outside,
tick pos=left,
x grid style={darkgray176},
xlabel={\(\displaystyle P_{J} (\si{\dBm})\)},
xmin=2, xmax=68,
xtick style={color=black},
y grid style={darkgray176},
ylabel={Rate (bits / s / Hz)},
ymin=-2.1129146308492, ymax=44.4567049981219,
ytick style={color=black}
]
\addplot [thick, black]
table {%
5 42.3399041058959
20 42.3399041058959
35 42.3399041058959
50 42.3399041058959
65 42.3399041058959
};
\addlegendentry{No Jammer}
\addplot [thick, black, dashed]
table {%
5 16.3861199234703
20 7.10035187861949
35 1.52049402451637
50 0.106716935019166
65 0.00388626137675397
};
\addlegendentry{No Protection}
\addplot [thick, blue, dash pattern=on 1pt off 3pt on 3pt off 3pt, mark=*, mark size=3, mark options={solid,rotate=180}]
table {%
5 35.2194256982448
20 29.556032999032
35 20.8114598743065
50 12.0643284116695
65 5.09275232843985
};
\addlegendentry{Analytic}
\addplot [thick, black, dashed, mark=x, mark size=3, mark options={solid,rotate=90}]
table {%
5 18.2024796239457
20 17.6870193396985
35 15.9886194195123
50 11.9966740098785
65 6.55537143840748
};
\addlegendentry{MinSINR}
\addplot [thick, blue, dash pattern=on 1pt off 3pt on 3pt off 3pt, mark=+, mark size=3, mark options={solid,rotate=90}]
table {%
5 26.2368879092419
20 26.1953056536591
35 26.1286608626437
50 25.3296954478561
65 21.1682137483008
};
\addlegendentry{ZF}
\end{axis}

\end{tikzpicture}

%% file: sections/figures/sum_rates_avg_aoa.tex
\begin{tikzpicture}

\definecolor{darkgray176}{RGB}{176,176,176}
\definecolor{lightgray204}{RGB}{204,204,204}

\begin{axis}[
legend cell align={left},
legend style={
  fill opacity=0.8,
  draw opacity=1,
  text opacity=1,
  at={(0.5, 1.1)},
  anchor=center,
  draw=lightgray204,
  font=\tiny,
},
legend columns=-1
tick align=outside,
tick pos=left,
x grid style={darkgray176},
xlabel={\(\displaystyle \theta_{J}\) (deg.)},
xmin=-38.25, xmax=33.25,
xtick style={color=black},
y grid style={darkgray176},
ylabel={Rate (bits / s / Hz)},
ymin=-1.11699520529479, ymax=44.4568993111907,
ytick style={color=black}
]
\path [draw=red, thick]
(axis cs:-10,0)
--(axis cs:-10,42);

\path [draw=red, thick]
(axis cs:0,0)
--(axis cs:0,42);

\path [draw=red, thick]
(axis cs:10,0)
--(axis cs:10,42);

\addplot [thick, black]
table {%
-35 42.3399041058959
-30 42.3399041058959
-25 42.3399041058959
-20 42.3399041058959
-15 42.3399041058959
-10 42.3399041058959
-5 42.3399041058959
0 42.3399041058959
5 42.3399041058959
10 42.3399041058959
15 42.3399041058959
20 42.3399041058959
25 42.3399041058959
30 42.3399041058959
};
\addlegendentry{No Jammer}
\addplot [thick, black, dashed]
table {%
-35 7.25127925382041
-30 6.86923188370905
-25 6.33067433068
-20 5.24785652278322
-15 4.32606944930112
-10 3.90952871788521
-5 3.18567470548896
0 3.20962153677538
5 3.21780951361517
10 3.88136948461482
15 4.31750527172573
20 5.31864291770682
25 6.3832756366961
30 6.88065403960384
};
\addlegendentry{No Protection}
\addplot [thick, blue, dash pattern=on 1pt off 3pt on 3pt off 3pt, mark=*, mark size=3, mark options={solid}]
table {%
-35 29.9586720330798
-30 28.5536855935162
-25 25.81254316876
-20 22.191505332371
-15 18.2090201175845
-10 14.346769497446
-5 13.3947594453311
0 12.7881688263615
5 13.3644370381041
10 14.3093678851877
15 18.1966362364853
20 22.1704931534361
25 25.8649227740119
30 28.5222169710641
};
\addlegendentry{Analytic}
\addplot [thick, black, dashed, mark=x, mark size=3, mark options={solid,rotate=270}]
table {%
-35 24.3264232434082
-30 21.7235681839057
-25 17.0338586598973
-20 14.5050714646505
-15 11.6653506966226
-10 7.69510907193026
-5 10.4959402108926
0 6.60735258658658
5 10.5314012706801
10 7.67735090723554
15 11.6633748078767
20 14.6032820111553
25 16.9739851502784
30 21.702390462919
};
\addlegendentry{MinSINR}
\addplot [thick, blue, dash pattern=on 1pt off 3pt on 3pt off 3pt, mark=+, mark size=3, mark options={solid}]
table {%
-35 40.2025452133025
-30 39.0215947871373
-25 36.1333225464539
-20 31.0455863687226
-15 19.5241968196675
-10 11.3254409855034
-5 12.5115024722304
0 10.6372710011535
5 12.4611774141774
10 11.2338918143243
15 19.6925963086434
20 31.0628580357289
25 36.1834683359134
30 39.1290860378058
};
\addlegendentry{ZF}
\draw (axis cs:-9.5,30) node[
  scale=0.4,
  anchor=base west,
  text=black,
  rotate=0.0
]{User 1};
\draw (axis cs:0.5,30) node[
  scale=0.4,
  anchor=base west,
  text=black,
  rotate=0.0
]{User 2};
\draw (axis cs:10.5,30) node[
  scale=0.4,
  anchor=base west,
  text=black,
  rotate=0.0
]{User 3};
\end{axis}

\end{tikzpicture}

%% file: sections/conclusions.tex
In this work, we derived a communication model which exploits the physical structure of the MIMO-MAC in the context of anti-jamming 
We derived one optimal strategy at the jammer, which seeks to minimize the rate of the strongest user.
We furthermore proposed an array of methods for protection against this worst-case jammer, 
possessing a far more powerful setup than the legitimate parties.
We have shown experimentally, that we can achieve a non-zero sum-rate even if the jamming and legitimate signals
come from roughly the same directions, and even if the jammer transmits with significantly more power.
We conclude, that resilience can not be fully guaranteed by only using the spatial dimension of communication.
Thus, our future work will be dedicated to the design of optimal signaling and frame structures, 
as well as to the study of optimal jamming strategies in these cases. 
